\documentclass[aps,prl,showpacs,superscriptaddress,floatfix,twocolumn,byrevtex]{revtex4-1}
\usepackage{amsmath}
\usepackage{amssymb}
\usepackage{amstext}
\usepackage{amsopn}
\usepackage{amsfonts}
\usepackage{amsxtra}
\usepackage{bm}
\usepackage{color}
\usepackage{dcolumn}
\usepackage{graphicx}
\usepackage{hyperref}
\begin{document}
\title{\boldmath $n_s - T_c$ \unboldmath Correlations in Granular Superconductors}
\author{Y. Imry}
\email{yoseph.imry@weizmann.ac.il}
\affiliation{Department of Condensed Matter Physics, Weizmann
 Institute of Science, Rehovot, 76100, Israel}
\author{M. Strongin}
\author{C. C. Homes}
\email{homes@bnl.gov}
\affiliation{Condensed Matter Physics and Materials Science Department,
 Brookhaven National Laboratory, Upton, New York 11973, USA}
\begin{abstract}
%
%
Following a short discussion of the granular model for an inhomogeneous superconductor,
we review the Uemura and Homes correlations  and show how both follow in two limits of a
simple granular superconductor model.  Definite expressions are given for the almost
universal coefficients appearing in these relationships in terms of known constants.

\end{abstract}
\date{\today}
%
%
%
\pacs{74.40.-n, 74.81.Fa \hfill
DOI:~\href{http://prl.aps.org/abstract/PRL/v109/i6/e067003}{10.1103/PhysRevLett.109.067003}
}
\maketitle

%
%
%
{\em Introduction}.---The model of a granular superconductor, in which superconducting
grains are coupled via Josephson tunneling, is considered \cite{deutscher73,*deutscher74a,
*emery95a,*emery95b,imry81,imry08,goldman88}.  This model is a very useful
paradigm both in its own right and because it is applicable to a number of real
situations. These range from man-made Josephson arrays to a variety of inhomogeneous
superconductors. Without insisting on  the relevance to high-$ T_c$ superconductors,
one may note that the ubiquitous phenomenon of the ``pseudogap'' in such
materials \cite{timusk99} finds a very natural qualitative explanation in this model.
This explanation is very clear in  the high intergrain resistance case
(see part A below).  The real transition  is the phase-locking of
the grains, while around the (higher) $T_c$ of the grain material, each grain
develops (continuously in temperature) a fluctuating order parameter which
leads to a smaller density of states near the Fermi level. Thus, the formation
of the pseudogap in this model is just a crossover, whose width is determined by,
and decreases with, the grain size.  These grains are evidently due to regions
in the system whose effective $T_c$ is higher due to fluctuations in the
doping level. The sensitivity of $ T_c$ to the doping level, compared with the
appropriate interface energy,  will determine whether, effectively, grains will
form (described later in the discussion). The case of a {\em d}-wave superconductor,
where $ T_c$ is sensitive to disorder, is immediately highlighted.

%
%
In this Letter we consider the question of the universal correlations reported
experimentally between the low-temperature superfluid density, $n_s$
and the transition temperature $T_c$ (Refs.~\onlinecite{uemura88,*uemura89,homes04,
homes05a,*homes05b,*homes09,*homes06,zuev05}).
Three such correlations have been reported for high-$T_c$
superconductors \cite{anderson97,dror}, and in some cases for
usual ``low-$T_c$'' ones. Two of them are different from each other,
while the third may be related to the second; these will be examined shortly.
It is of great interest to understand the physics behind
such correlations \cite{zaanen04,*phillips05,maple07} and their
respective ranges of validity \cite{tallon05,schneider05,rameau11}.
We show that both these clearly different correlations
follow from two limits of a simple classical granular superconductor
model \cite{deutscher73,*deutscher74a,*emery95a,*emery95b,imry81,imry08,goldman88}
and derive the coefficients in terms of natural constants and the gap-to-$T_c$
ratio for the underlying grain material. Our derivation is embarrassingly simple.
The Uemura/Homes law follows when the critical temperature for the
intergrain phase locking is much smaller/comparable to that for the
grain material. The strongly inhomogeneous, granular, picture is a
broadly applicable paradigm \cite{dag}, describing many diverse
systems \cite{imry08,strongin70,*deutscher74b,*simon87,*shahar92,*ekinci98,
*merchant01,kowal94}.  Recently, there is strong evidence for the relevance
of this paradigm also for high temperature superconductors \cite{tranquada95,*howald01,*pan01,*lang02,*mcelroy05,*fang05,
*gomes07,*wise09,*append}.

In 1988, Uemura {\em et al.}~\cite{uemura88,*uemura89} reported, for underdoped
high-$T_c$ superconductors, the proportionality of $n_s/m^*$ (or
$\lambda^{-2}$, where $\lambda$ is the penetration length and $m^*$
the effective carrier mass, which is of the order of $5m$ for most of
the considered materials, where $m$ is the electron mass)
to $T_c$. Here $n_s$ was determined from the muon spin relaxation rate
for four high $T_c$ families with varying doping level (carrier density).
The coefficient in the linear relationship is such that a carrier
density of $n_s = 2\times 10^{21}$~cm$^{-3}$ corresponds to $T_c\simeq 25$~K.

\begin{figure}[t]
\includegraphics[width=1.0\columnwidth]{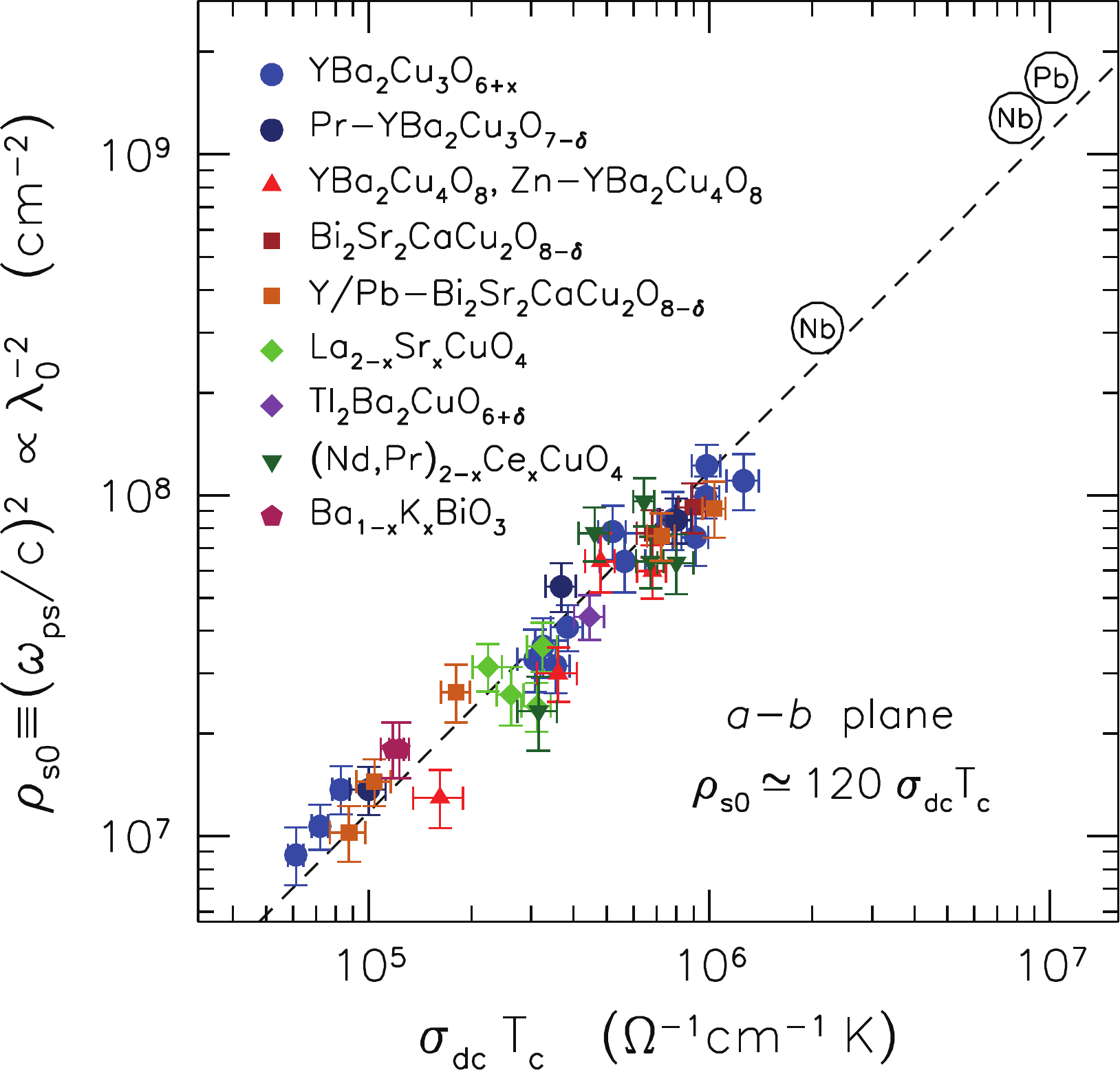}
\caption{A log-log plot of the superfluid density $\rho_{s0}$
 vs $\sigma_{dc}\,T_c$ in the {\em a-b} planes for a variety of
 electron and hole-doped cuprates.
 The dashed line corresponds to the general result for the cuprates
 $\rho_{s0} \simeq 120\, \sigma_{dc} T_c$.  The points for Nb and Pb,
 indicated by their atomic symbols, also fall close to this line.}
\label{fig:scale}
\end{figure}

In 2004, Homes {\em et al.}~\cite{homes04} reported a different correlation,
valid more generally, including the overdoped and optimally doped
cases: $\rho_{s0} \simeq 120\, \sigma_{dc} T_c$, where $\rho_{s0}$
is the strength of the condensate  determined by optical measurements,
and $\sigma_{dc}$ is the normal-state dc conductivity near $T_c$.  The
superfluid density is related to the superconducting plasma
frequency $\rho_{s0} \equiv  \omega_{ps}^2 \propto n_s/m^*$ as well
as the penetration depth, $\rho_{s0} = c^2/\lambda_0^2$.
Nine different high-$T_c$ material families with varying doping
(including optimal and beyond) were examined, as well as the
phonon-mediated superconductors Pb and Nb, shown in Fig.~\ref{fig:scale}.
This result has been interpreted \cite{homes05a,*homes05b,*homes09,*homes06} in terms of the
conventional decrease of $n_s$ proportional to $\ell/ \xi_0 \propto T_c \tau $
in the dirty limit of BCS superconductors, where $\tau$ and $\ell$ are the
mean-free time and scattering length and $\xi_0$ the zero-temperature BCS
coherence length ($\xi_0 \propto v_F/T_c$). The questions of why
these materials are in the dirty limit, when $T_c$ is so high and
the coherence length so small, and to what extent can the BCS-type
relationships be used for high $T_c$ materials (in spite of current
theoretical beliefs) were left open. Clearly, the $d$-wave nature of
these superconductors might play an important role here.

Finally, in 2005 Zuev {\em et al.}~\cite{zuev05} reported a linear
relationship between $n_s$ and $T_c^{\chi}$, where $\chi = {2.3
\pm 0.4}$. They pointed out that with the empirical
proportionality of $T_c$ to $\sigma_{dc}$ (theoretically justified
in a large parameter regime for a classical Josephson-coupled
superconductor \cite{imry81}, see below); a value of $\chi=2$ is
within the experimentally-determined range and would make their
result consistent with the one by Homes {\em et al.}~\cite{homes04}.

These quite universal correlations have caused considerable
discussion \cite{zaanen04}.  For a recent explanation, we mention
the one relying on the vortex glass melting temperature \cite{maple07}.

%
%

%
%
{\em Method.}---We shall now demonstrate that the various $n_s \propto T_c$
relations follow in an almost trivial manner for a classical (no
capacitive energies) ordered Josephson array \cite{imry08}. We take
the simplest model of a two-dimensional (2D) array of square $L\times L$
grains of thickness $d$ in the {\em x--y} plane, made from a
superconductor with a critical temperature $T_c^{0}$. The grains are
connected by flat Josephson junctions with Josephson current
amplitudes $I_J$ and Josephson energies $E_J = \hbar I_J / 2e$.
The 2D array can be regarded as the whole system or as one of the
layers in a 3D structure. From now on we mainly consider ``large
grains'' in the Anderson sense \cite{anderson}. There, the intragrain
gap is much larger than the single-particle level spacing, $w_L$,
of the isolated grain. In such ``large'' grains, bulk
superconductivity is approximately valid. The Josephson coupling
can be written as \cite{jj}
\begin{equation}
  E_J = (\pi/4) g_n  \Delta,
  \label{Jos}
\end{equation}
where $g_n$ is the intergrain conductance measured in units of
$e^2 / \hbar$.  We assume, for definiteness, that the size, $L$,
of each superconducting unit is $\ll \lambda_0$, where $\lambda_0$
is the penetration depth of the grain material. (It is straightforward
to get the result for $L \gg \lambda_0$, as well). This immediately
implies that $L$ is much smaller than the effective penetration depth
for the array, i.e.~all induced fields are neglected.  This can be taken as a
model for a granular superconductor as long as the effects of the capacitances
and the intergrain disorder, which certainly exist in real cases, are not
dominant \cite{giaever68,*abeles78,*kawabata77,*yurkevich01,*ambegaokar82}.

We now obtain the linear response to a small magnetic field $B_z$
perpendicular to the array. For $\lambda_0 \gg L$ the field $B_z$ is
uniform over each grain. $\vec B$ is derived from a vector potential
$\vec A = (B_{z}y, 0, 0)$. Note that $\nabla\cdot \vec{A} = 0$ as required
for the London gauge. Thus the London equation takes the form
\begin{equation}
  j_s = -\frac{n_s e^2}{m^* c} A.
  \label{London}
\end{equation}

The flux enclosed in an $L\times y$ rectangle shared equally by two neighboring
grains is ${B_z}Ly$. Due to it, the phase difference between two superconducting
blocks that are nearest neighbors in the $x$ direction, increases
with $y$ in the manner
\begin{equation}
  \phi(y) \simeq -2eB_zyL/\hbar c= -2eL A_x(y)/\hbar c.
\label{phase}
\end{equation}
For small $B$, this leads to a Josephson current density \cite{jj}
\begin{equation}
  j_{s,x}(y) = -2eI_J A_x(y)/\hbar c d.
\label{current}
\end{equation}
Equating this Josephson current to the screening current in the London equation
[Eq.~(\ref{London})], we find the general relation for a granular superconductor
with $L \ll \lambda_0$ is similar in form to the result for an array of
superconducting weak links \cite{lobb83}
\begin{equation}
  n_{s} = \frac{4m^*}{d\hbar^2 } E_J.
\label{ns}
\end{equation}
\noindent This relation can be written in terms of $T_c$ and the
normal-state conductivity. Different results are obtained in the
two following cases.

%
%
%
%
{\em A. Large intergrain resistance $g_n \ll 1$.}---
Because the electrons are well localized in the grains, one expects the
normal state of this system to be insulating when extrapolated to
$T\rightarrow 0$ \cite{imry08}. Here, to reach $E_J \sim T$, one needs
 to go to temperatures much lower than $T_c^{0}$ [see Eq.~(\ref{Jos})].
At those temperatures $E_J$ saturates with values of order $g_n
\Delta(0)$, and $T_c$ is given by a constant $\zeta$ of order
unity times $E_J$, $T_c = \zeta E_J$.  We have used units in
which $k_B = 1$ throughout. Thus, in this case we obtain
\begin{equation}
 n_{s} = \frac{4m^*}{d\hbar^2 \zeta} T_c,
\label{Uemura}
\end{equation}
which is just the Uemura correlation.
For $m^{*} = 5m, \zeta = 1, d = 5$~\AA\ and $n_s = 2\times 10^{21}$~cm$^{-3}$
we obtain $T_c \approx 35$~K. Thus, the coefficient in Eq.~(\ref{ns})
agrees within a factor of two with the Uemura one, for
reasonable parameters of the 2D layer. Eq.~(\ref{ns}) is just the
relation between $n_s$ and the order-parameter phase stiffness for
the $XY$ model. In this limit, there exist two distinct effects,
the buildup of the pairing correlations in the grains, which is a
continuous crossover around $T_c^0$, and the intergrain phase
locking at $T_c$.  Were this picture applicable to the underdoped
high-$T_c$ case, $T_c^0$ and $T_c$ would correspond to the
establishment of the pseudogap and that of overall superconductivity
respectively.

When the Uemura correlation was first reported, the
proportionality of $T_c$ to the  electron density was taken to
indicate the purely electronic origin of high-$T_c$
superconductivity. Our simple derivation above proves that that
logic is not infallible. The Josephson array can model any
appropriately inhomogeneous superconductor, including ordinary
low-$T_c$ ones, and it does yield the Uemura correlation.

%
%
%
{\em B. Small intergrain resistance $g_n \gtrsim 1$, (including $g_n \gg 1$).}---
Here $E_J$ becomes comparable to $T$ around $T_c^{0}$, which
is then approximately equal to $T_c$. In the high $T_c$ case, this
would mean that the pseudogap and the superconductivity are
established at the same temperature, which is the case for the
optimally doped and overdoped situations.  Here we obtain, defining
the constant $A$ ($\cong 1.75$ in the BCS case) via the usual
relationship in the grain material, $\Delta(0)= A T_c^0$:
\begin{equation}
  n_{s} = \frac{A \pi g_n m^*}{d\hbar^2 }\,T_c
        = \frac{A \pi \sigma^\prime_n m^*}{\hbar^2 }\,T_c,
\end{equation}
which is equivalent to the Homes' law! Here $\sigma^\prime_n$ is
the normal state conductivity just above $T_c$, in units of
$(e^2/\hbar)/{\rm cm}$.  It makes sense that the condition $g_n
\gtrsim 1$ is in fact satisfied for high enough doping.  While
the square of the superconducting plasma frequency is defined
as $\omega_{ps}^2 = n_s e^2/(\pi m^*)$ in units s$^{-2}$, the
quantity that Homes {\em et al.} examines is $(\omega_{ps}/c)^2$
in units of cm$^{-2}$; this allows us to cast the Homes' law in the
following seemingly  elegant way.  Using the ``universal''
constant $\alpha^\prime  \equiv  A \alpha \simeq 0.0128$,
where $\alpha \equiv e^2/ (\hbar c)$ is the fine structure
constant and we have previously taken $A = 1.75$:
\begin{equation}
  (\omega_{ps}/c)^2 = \pi\alpha^\prime \sigma^\prime_n T_c /(\hbar c).
\label{univ}
\end{equation}

\noindent The approximate coefficient in the proportionality of
the LHS to $\sigma_n^\prime T_c$ depends {\em only on the gap-to-$T_c$
ratio and on natural constants}.  We note that this result agrees
exactly with the usual ``dirty limit'', in which the reduction of
$n_s$ is: $n_s = n\, (v_{\rm F} \tau/\xi_0)$.  Using the BCS-type
relationship $\xi_0 = \hbar v_{\rm F}/[\pi \Delta(0)]$ yields Eq.~(\ref{univ}).
Since the dirty-limit value agrees approximately with the Homes' law
coefficient \cite{homes04,homes05a,*homes05b,*homes09,*homes06}, so should
our result [Eq.~(\ref{univ})].

In fact, since $e^2/\hbar \simeq 1/4100$~$\Omega^{-1}$, $\sigma_n^\prime
\backsimeq 4100\,\sigma_n$, when $\sigma_n$ is measured in
$\Omega^{-1}{\rm cm}^{-1}$. This yields
\begin{equation}
  (\omega_{ps}/c)^2 \backsimeq 50\, \sigma_n T_c /(\hbar c)
  \backsimeq 235\, \sigma_n T_c .
  \label{homes}
\end{equation}
In the last expression on the RHS $\sigma_n$ and $T_c$ are in the same units
as in the original Homes law paper \cite{homes04} (namely $\Omega^{-1}{\rm cm}^{-1}$
and K) \cite{scaling}.  The slope in Eq.~(\ref{homes}) is larger by about a factor
of two from the value $120\pm 25$ reported in Ref.~\onlinecite{homes04}.  Note that
this expression can be recast as $T_c \propto \rho_{s0}\rho_{dc}$, where $\rho_{dc}
= 1/\sigma_n$, allowing a more direct comparison with the Uemura relation.
For the low-$T_c$ cases, the agreement is within about 50 percent.
Taking, as an example, the first Nb sample of Table~I of
Ref.~\onlinecite{homes05b}, we get from the values of $\sigma_n$ and
$T_c$, $(\omega_{ps}/c)^2 \sim 4.7\times 10^8$~cm$^{-2}$, while the
value in the table is $\sim 3.1\times 10^8$~cm$^{-2}$.
Overall, the agreement of our Eq.~(\ref{homes}) with experiment is
within a factor of two. With the gross simplifications introduced
in our na\"{\i}ve model, we regard this as satisfactory.

%
%
%
{\em Discussion}.---
Nominally, the high $T_c$ superconductors look like they are in the
clean limit. This is believed rather generally and is consistent with
the values reported in Refs.~\onlinecite{homes04,homes05a,*homes05b,*homes06}.
This is not because they are so clean; the experimental values of
$k_F \ell$ can be of order $50-100$. The problem is that $\xi_0$ is small
due to the relatively large ratio $T_c / E_F$ (here $k_F \xi_0 \sim 10-50$).
Thus, these materials would be in the clean limit {\em if they were
homogenous}. However, the natural fluctuations in the doping make
them inhomogeneous, and $\xi_0$ is not large enough to average that.
What is shown above is that {\em this inhomogeneous system behaves
like a dirty one}. (Although, again, it would not be,  were it
homogenous!)
How the inhomogeneity arises has been discussed by Alvarez
and Dagotto \cite{dag} and more recently by Hoffman \cite{jhnote};
we elaborate on this in Ref.~\onlinecite{imry08} following the general
argument by Ma and one of us \cite{IM}.
The issue at hand is
to find the size scale where the gain in energy, $\propto L^{d/2}$
due to fluctuations in the defect concentration or doping is larger
than the energy of the interface created, $\propto L^{d-2}$.
When this scale of $L$ is smaller than the effective correlation
length, the instability and formation of grains will occur. Clearly,
a stronger sensitivity of $T_c$ to the defect concentration will help
in the establishment of the granular state of matter.  As mentioned
before, a $d$-wave character of the superconductor is very helpful in
that respect.


The behavior of the granular model presented is determined by two
conditions, governed by dimensionless ratios: whether the grains
are large or small [depending on the ratio $\Delta(0)/w_L$] and
whether the intergrain resistance (in units of $\hbar/e^2$) is
larger or smaller than unity.
%
%
The Uemura correlation is valid only for the large grain, large
resistance case which typically corresponds to the underdoped
region of the phase diagram for the cuprates where a pseudogap is
usually observed \cite{timusk99}; $\Delta(0)/w_L \gtrsim 1, g_n \ll 1$,
where the insulator and the inhomogeneous Josephson phase are the
relevant phases.
%
%
In the large grain, small resistance case (which corresponds to the
optimally or overdoped cuprates where the pseudogap is either strongly
diminished or absent altogether), $\Delta(0)/w_L \gtrsim 1, g_n \gg 1$,
we showed that the Homes' law is the relevant correlation.
%
%
In the small grain case $\Delta(0)/w_L <1$ and small resistance
metallic regime $g_n \gtrsim 1$ (the optimal and overdoped regime),
superconductivity is established in an almost homogenous, strongly
disordered, conductor.  Even close to the metal-insulator transition
the mean free path $\ell$ is of a small microscopic size and it
makes sense that the superconductor should be in the dirty limit
($\ell \ll \xi_{0}$).  This implies that the Homes' law
\cite{homes04,homes05a,*homes05b,*homes09,*homes06} (or the one
reported by Zuev {\em et al.}~\cite{zuev05}) should then yield the valid
correlation between $n_s$ and $T_c$.
%
%
The small grain case $\Delta(0)/w_L \lesssim 1$ with large intergrain
resistance $g_n \lesssim 1$ (and therefore an insulating normal state)
is very interesting since strong intragrain superconducting correlations
do exist \cite{Moshe}, but it is not known exactly what is the effect of
the intergrain coupling with the Coulomb blockade \cite{giaever68,*abeles78,
*kawabata77,*yurkevich01,*ambegaokar82}.

We have neglected throughout the capacitive, Coulomb-blockade-type
interactions \cite{giaever68,*abeles78,*kawabata77,*yurkevich01,*ambegaokar82}.
This is justified in all the metallic regimes, due to screening.
This includes all the range of the Homes correlations. The capacitive
effects might also play an important role in the
large-grain-small-intergrain-coupling case, where the
Uemura correlations are supposed to hold. Our treatment there is
valid only for grains large enough for the capacitive effects not
to be important.

Only compact regular grains were considered in this paper. More
general inhomogeneities (e.g. stripes \cite{dror}, layers or more
complex geometries) should be treated as well. The case of high
$T_c$ materials is further complicated due to the anisotropic gap
and correlation length \cite{jog}. The question of when such a
superconductor can be regarded as dirty is beyond the scope of this
paper.

The results presented in this paper are valid for granular
superconductors and Josephson arrays. They do explain
semiquantitatively the $n_s - T_c$ correlations in ordinary and in
high $T_c$ superconductors. This obviously does {\em not prove} that
the latter conform to the na\"{\i}ve model discussed. However, it might
be taken as further evidence that the inhomogeneities, which do
exist \cite{tranquada95,*howald01,*pan01,*lang02,*mcelroy05,*fang05,
*gomes07,*wise09,*append}
in these materials, play a role in their fascinating physics.

%
%
{\em Conclusions.}---The examination of the model for a granular, inhomogeneous
superconductor reveals that it can mimic a dirty-limit material.  Scaling relations
may be derived in the two limits of a simple granular superconductor model.
In the large grain, high resistivity case [$\Delta(0)/\omega_L \gtrsim 1,\, g_n\ll 1$]
Uemura-type $T_c \propto n_s$ (or $T_c \propto \rho_{s0}$) scaling is expected; however,
in the small resistance case [$\Delta(0)/\omega_L \gtrsim 1,\, g_n\gg 1$] it is demonstrated
that the $T_c \propto \rho_{s0}\rho_{dc}$ scaling described by Homes {\em et al.} is expected.
In the small grain, small resistance case [$\Delta(0)/\omega_L < 1,\, g_n\gg 1$],
$T_c \propto \rho_{s0}\rho_{dc}$ is again expected.  Within the context of the
high-temperature superconductors, the high-resistivity case corresponds to the
underdoped (pseudogap) region of the phase diagram where Uemura-type scaling is
observed, while the low-resistivity case corresponds to the optimally and overdoped
region where the pseudogap is largely absent and $T_c \propto \rho_{s0}\rho_{dc}$
is observed rather than $T_c \propto \rho_{s0}$.

%
%
Y.~I.~thanks Ehud Altman, Alexander Finkelstein, Amit Keren and Zvi Ovadyahu
for helpful discussions and comments; C.~C.~H.~wishes to thank J.~E.~Hoffman
and J.~D.~Rameau for useful discussions.
The work of Y.~I. was supported by a Center of Excellence
of the Israel Science Foundation (ISF, grant No.~1566/04), and by
the German Federal Ministry of Education and Research (BMBF) within
the framework of the German-Israeli project cooperation (DIP).  Work at
Brookhaven was supported by the U.S. Department of Energy, Office of
Basic Energy Sciences, Division of Materials Sciences and Engineering
under Contract No. DE-AC02-98CH10886.
%
%
%
%
%

\begin{thebibliography}{54}
\bibitem [{app()}]{append}%
  \BibitemOpen
  \href@noop {} {}\bibinfo {note} {This work is a selection showing
  inhomogeneities in high-temperature superconductors. Note that in this work
  the term ``granular'' is used in a general sense and this term includes
  electronic inhomogeneities as well as the traditional structural
  granularity.}\BibitemShut {Stop}%
\bibitem [{sca()}]{scaling}%
  \BibitemOpen
  \href@noop {} {}\bibinfo {note} {The scaling relation can be presented with
  both $(\omega_{ps}/c)^2$ and $\sigma_{dc}\,T_c$ in units of cm$^{-2}$; this
  is accomplished by noting that $1\,\Omega^{-1}{\rm cm}^{-1} \simeq
  4.76$~cm$^{-1}$ and $1\,{\rm K} \simeq 0.69$~cm$^{-1}$. In this
  representation, the calculated slope of the spectral weight of the condensate
  $\rho_{s0}/8 \simeq 8.1\,\sigma_{dc} T_c$, which is roughly twice the
  observed value of $\rho_{s0}/8 \simeq 4.4\,\sigma_{dc}T_c$.}\BibitemShut
  {Stop}%
\end{thebibliography}
%
\providecommand{\noopsort}[1]{}\providecommand{\singleletter}[1]{#1}

\end{document}